\documentclass{emulateapj}

\slugcomment{Accepted by ApJ}
\shorttitle{A Bipolar Supernova Remnant in the SMC}
\shortauthors{LOPEZ ET AL.}

\newcommand{\ltsima}{$\; \buildrel < \over \sim \;$}
\newcommand{\simlt}{\lower.5ex\hbox{\ltsima}}

\newcommand{\ls}{{_<\atop^{\sim}}}

\newcommand{\gs}{{_>\atop^{\sim}}}

\newcommand{\source}{SNR~0104$-$72.3}

\begin{document}

\title{Identification of a Jet-Driven Supernova Remnant in the Small Magellanic Cloud: Possible Evidence for the Enhancement of Bipolar Explosions at Low Metallicity}

\author{Laura A. Lopez\altaffilmark{1,5,6}, Daniel Castro\altaffilmark{1}, Patrick O. Slane\altaffilmark{2}, Enrico Ramirez-Ruiz\altaffilmark{3}, Carles Badenes\altaffilmark{4}}
\altaffiltext{1}{MIT-Kavli Institute for Astrophysics and Space Research, 77 Massachusetts Avenue, 37-664H, Cambridge MA 02139, USA}
\altaffiltext{2}{Harvard-Smithsonian Center for Astrophysics, 60 Garden St., Cambridge, MA 02138, USA}
\altaffiltext{3}{Department of Astronomy and Astrophysics, University of California Santa Cruz, 1156 High Street, Santa Cruz, CA 95060, USA}
\altaffiltext{4}{Department of Physics and Astronomy and Pittsburgh Particle Physics, Astrophysics and Cosmology Center (PITT PACC), University of Pittsburgh, 3941 O'Hara St, Pittsburgh, PA 15260, USA}
\altaffiltext{5}{NASA Einstein Fellow}
\altaffiltext{6}{Pappalardo Fellow in Physics}

\email{lopez@space.mit.edu}

\begin{abstract}

Recent evidence has suggested that the supernova remnant (SNR) 0104$-$72.3 in the Small Magellanic Cloud (SMC) may be the result of a ``prompt'' Type Ia SN based on enhanced iron abundances and its association with a star-forming region. In this paper, we present evidence that \source\ arose from a jet-driven bipolar core-collapse SN. Specifically, we use serendipitous {\it Chandra X-ray Observatory} data of \source\ taken due to its proximity to the calibration source SNR~E0102$-$72.3. We analyze 56 Advanced CCD Imaging Spectrometer (ACIS) observations of \source\ to produce imaging and spectra with an effective exposure of 528.6 ks. We demonstrate that \source\ is highly elliptical relative to other nearby young SNRs, suggesting a core-collapse SN origin. Furthermore, we compare ejecta abundances derived from spectral fits to nucleosynthetic yields of Type Ia and core-collapse (CC) SNe, and we find that the iron, neon, and silicon abundances are consistent with either a spherical CC SN of a 18--20 $M_{\sun}$ progenitor or an aspherical CC SN of a 25 $M_{\sun}$ progenitor. We show that the star-formation history at the site of \source\ is also consistent with a CC origin. Given the bipolar morphology of the SNR, we favor the aspherical CC SN scenario.  This result may suggest jet-driven SNe occur frequently in the low-metallicity environment of the SMC, consistent with the observational and theoretical work on broad-line Type Ic SNe and long-duration gamma-ray bursts. 

\end{abstract}


\section{Introduction}

Hundreds of supernovae (SNe) are discovered each year at optical wavelengths by dedicated surveys which focus on nearby galaxies (e.g., \citealt{law05,leaman11}). Despite their high frequency, these SNe are too distant to resolve the SN ejecta and the immediate surroundings of the exploded stars. Fortunately, supernova remnants (SNRs) offer a complementary means to study SN explosions and dynamics up close and in detail. 

Currently, $\sim$350 SNRs have been identified in the Milky Way (MW) and the Large and Small Magellanic Clouds (LMC and SMC, respectively) through multiwavelength campaigns (e.g., \citealt{green09,bad10}). The morphologies and spectral characteristics of these SNRs provide crucial insights regarding the nature of SN explosions, the environments of progenitors, and the interactions of massive stars and their circumstellar media (e.g., \citealt{bad07,lopez09b,lopez11,vink12}). In particular, study of SNRs will aid in addressing two outstanding issues in the literature: the progenitors of Type Ia SNe and the triggering mechanism of jet-driven explosions. 

Regarding the former topic, it is currently debated whether the progenitor of Type Ia SNe is a single white dwarf accreting from a non-degenerate companion (e.g., \citealt{whelan73}) or the merger of a double white-dwarf system (e.g. \citealt{iben84,webbink84,james,dan12,pakmor12}). Nearby Type Ia SNRs can be investigated for clues regarding the nature of the progenitors, including searches for a surviving non-degenerate companion (e.g., \citealt{schaefer12,kerz13}) and evidence of a circumstellar medium modified by a non-degenerate companion (e.g., \citealt{williams11}). 

Secondly, many mysteries remain regarding the nature of jet-driven SNe. Jet-driven core-collapse (CC) SNe are thought to occur among $\sim$1--2\% of Type Ib/c SNe \citep{soderberg10}, and some fraction of these events are associated with long gamma-ray bursts (GRBs; e.g., \citealt{izzard04,pod04}). As GRBs are typically detected at cosmological distances (see review by \citealt{gehrels09}), identification of local analogues among the nearby SNR population would help constrain the physics, dynamics, and nucleosynthesis of these explosions (e.g., \citealt{ramirezruiz10,lopez13b,lopez13,diego13}).

Recently, it was reported that \source\ in the SMC may be the remnant of an unusual Type Ia explosion \citep{lee11}. In particular, these authors showed that modeling of X-ray spectra from \source\ required an overabundance of iron, consistent with a Type Ia origin for this SNR. As this source is also associated with a nearby star-forming region (e.g., \citealt{koo07}), Lee et al. concluded that the progenitor of \source\ may have been a white dwarf from a young population of stars, i.e. from a ``prompt'' Type Ia SN (e.g., \citealt{scann05,aubourg08}).

To further investigate the unusual nature of this SNR, in this paper we take advantage of serendipitous observations of \source\ with the {\it Chandra X-ray Observatory}. Due to its close proximity ($\sim$11\arcmin) to the {\it Chandra} calibration source SNR~E0102$-$72.3, \source\ has been observed regularly throughout the lifetime of {\it Chandra}. When combined, these data are nearly five times deeper than the targeted {\it Chandra} observations presented by \cite{lee11}. Using these data, we present evidence that the SNR actually arose from a jet-driven CC SN. In Section~\ref{sec:data}, we detail the archival observations of \source\ and how we merged these data. In Section~\ref{sec:results}, we utilize these combined data to explore the explosive origin of \source, based on its morphology (Section~\ref{sec:morphology}) and its metal abundances (Section~\ref{sec:spectra}). We present a discussion and conclusions in Section~\ref{sec:conclusions}, with emphasis on the implication of finding a jet-driven CC SNR in the low-metallicity SMC.

\section{Data Reduction} \label{sec:data}

We identified the available {\it Chandra} imaging of \source\ using the following procedure. First, we searched the {\it Chandra} archive for ACIS-I and \hbox{ACIS-S} observations within 20\arcmin\ of \source\ without gratings and in timed-exposure mode. This query yielded 242 observations: 237 calibration observations of SNR~E0102$-$72.3, as well as pointed observations of \source\ (ObsIDs 9100 and 9810), of the nearby SNR~0103$-$72.6 (ObsID 2758), and of the SMC wing (ObsIDs 5485 and 5486). Then, we visually inspected the 242 observations to identify all those where the entirety of \source\ was imaged by an ACIS chip, regardless of off-axis distance. Using this strategy, we found 56 ACIS-I and ACIS-S observations of \source\ from 2000--2009, with a combined integration of 528.6~ks. These observations are listed in Table~\ref{table:obs} along with relevant details for each pointing, including the ACIS chip on which \source\ is located, the off-axis distance of \source\ to the aim point, and the net (background-subtracted) full-band (0.5--8.0 keV) counts obtained. The median off-axis distance of \source\ in the 56 observations is 10\arcmin. For reference, at 1.49 keV, the point-spread function\footnote{Defined as the radius where 50\% of encircled energy of a point source is detected: http://cxc.harvard.edu/proposer/POG/html/index.html} is $\approx$6\arcsec at 10\arcmin\ off axis, small enough to resolve substructures in the $\sim$1.7\arcmin\ diameter \source. In the analyses described below, we use the software {\it Chandra} Interactive Analysis of Observations ({\sc ciao}) Version 4.3 and XSPEC Version 12.7.0 \citep{arnaud96}.

\begin{deluxetable}{llrcrr}
\tablecolumns{6}\tablewidth{0pc} \tabletypesize{\footnotesize}
\tablewidth{0pt} \tablecaption{Observations of \source, Sorted by ObsID} 
\tablehead{\colhead{ObsID} & \colhead{Obs Date} & \colhead{Exposure} & \colhead{Chip\tablenotemark{a}} & \colhead{Off-Axis} & \colhead{Net} \\
\colhead{} & \colhead{} & \colhead{(ks)} & \colhead{} & \colhead{Distance\tablenotemark{b}} & \colhead{Counts\tablenotemark{c}}}
\startdata
1316 & 2000-12-15 & 6.87 & S2 & 12.4 & 358 \\
1317 & 2000-12-15 & 6.87 & S2 & 14.4 & 357 \\
1528 & 2000-12-16 & 6.87 & I2 & 5.1 & 303 \\
1544 & 2001-06-05 & 7.42 & I1 & 5.6 & 398 \\
2835 & 2001-12-05 & 7.83 & I2 & 6.6 & 365 \\
2836 & 2001-12-05 & 7.46 & I2 & 8.0 & 342 \\
2839 & 2001-12-05 & 7.46 & S2 & 13.3 & 337 \\
2841 & 2001-12-06 & 7.46 & I2 & 5.2 & 410 \\
2850 & 2002-06-19 & 7.76 & I3 & 10.1 & 321 \\
2852 & 2002-06-19 & 7.56 & I2 & 10.3 & 298 \\ 
2864 & 2002-06-21 & 7.56 & I1 & 6.0 & 366 \\
3529 & 2003-08-09 & 7.65 & I1 & 8.4 & 259 \\
3530 & 2003-08-09 & 7.65 & I1 & 6.6 & 335 \\
3531 & 2003-08-10 & 8.07 & I1 & 4.7 & 314 \\
3534 & 2003-08-10 & 7.66 & I0 & 10.1 & 256 \\
3536 & 2003-02-02 & 7.63 & S3 & 12.3 & 576 \\
3537 & 2003-02-02 & 7.63 & S3 & 11.4 & 565 \\
3538 & 2003-02-02 & 7.63 & S3 & 14.2 & 551 \\
3539 & 2003-02-02 & 7.63 & S3 & 16.1 & 551 \\
3540 & 2003-02-02 & 7.63 & S3 & 18.0 & 497 \\
3541 & 2003-02-02 & 7.63 & I2 & 6.2 & 335 \\
3542 & 2003-02-02 & 7.63 & I3 & 8.8 & 315 \\ 
3543 & 2003-02-02 & 7.64 & S2 & 13.7 & 390 \\ 
3544 & 2003-08-10 & 7.86 & I2 & 10.7 & 249 \\ 
3547 & 2003-08-08 & 7.66 & I2 & 12.3 & 260 \\ 
5130 & 2004-04-09 & 19.41 & S4 & 10.0 & 891 \\ 
5131 & 2004-04-05 & 8.01 & S4 & 11.9 & 134 \\ 
5132 & 2004-04-09 & 7.50 & S2 & 4.1 & 431 \\ 
5133 & 2004-04-09 & 7.50 & S1 & 11.3 & 547 \\ 
5134 & 2004-04-09 & 7.50 & S3 & 6.2 & 611 \\ 
5135 & 2004-04-10 & 8.14 & S5 & 19.9 & 271 \\ 
5143 & 2004-04-26 & 7.35 & I1 & 8.2 & 308 \\ 
5145 & 2004-04-28 & 7.15 & I3 & 6.8 & 353 \\ 
5150 & 2003-12-19 & 7.57 & S2 & 12.7 & 353 \\ 
5151 & 2003-12-19 & 7.57 & S2 & 14.7 & 305 \\
5153 & 2003-12-16 & 7.40 & I2 & 5.1 & 254 \\ 
6042 & 2005-04-12 & 18.90 & S4 & 10.0 & 858 \\ 
6043 & 2005-04-18 & 7.85 & S4 & 11.9 & 344 \\ 
6049 & 2004-12-13 & 7.58 & I2 & 7.6 & 272 \\ 
6050 & 2004-12-13 & 7.16 & I2 & 8.5 & 260 \\ 
6051 & 2005-01-12 & 17.92 & S2 & 11.1 & 630 \\ 
6053 & 2005-01-12 & 7.16 & S3 & 14.8 & 460 \\ 
6054 & 2005-01-12 & 7.17 & S3 & 16.8 & 447 \\ 
6056 & 2004-12-17 & 8.01 & I2 & 5.2 & 224 \\ 
6753 & 2006-03-14 & 7.17 & I3 & 3.8 & 298 \\
8361 & 2007-02-05 & 19.79 & S3 & 11.4 & 1405 \\ 
8362 & 2007-02-11 & 8.84 & I2 & 5.1 & 391\\ 
8363 & 2007-02-11 & 8.45 & I3 & 9.7 & 340  \\ 
8364 & 2007-02-11 & 8.45 & S2 & 13.0 & 248 \\
9100 & 2008-01-27 & 54.47 & S3 & 0.6 & 4762 \\ 
9691 & 2008-02-04 & 7.94 & I2 & 5.9 & 340 \\ 
9692 & 2008-02-05 & 7.67 & I3 & 9.0 & 245 \\ 
9693 & 2008-02-05 & 7.68 & S2 & 13.5 & 335 \\ 
9810 & 2008-01-30 & 55.82 & S3 & 0.6 & 4745 \\ 
10650 & 2009-02-16 & 7.93 & I2 & 4.4 & 350 \\ 
10652 & 2009-01-17 & 8.07 & S2 & 14.7 & 259 \\
\enddata
\tablenotetext{a}{ACIS chip where \source\ is located in the observation.} 
\tablenotetext{b}{Angular distance in arcminutes from the observation aim point to the position of \source, at right ascension 01h06m14s and declination $-$72d05m18s (J2000).}
\tablenotetext{c}{Total number of background-subtracted source counts in the 0.5--8.0 keV band during the observation.}
\label{table:obs}
\end{deluxetable}

\begin{figure*}
\begin{center}
\includegraphics[width=\textwidth]{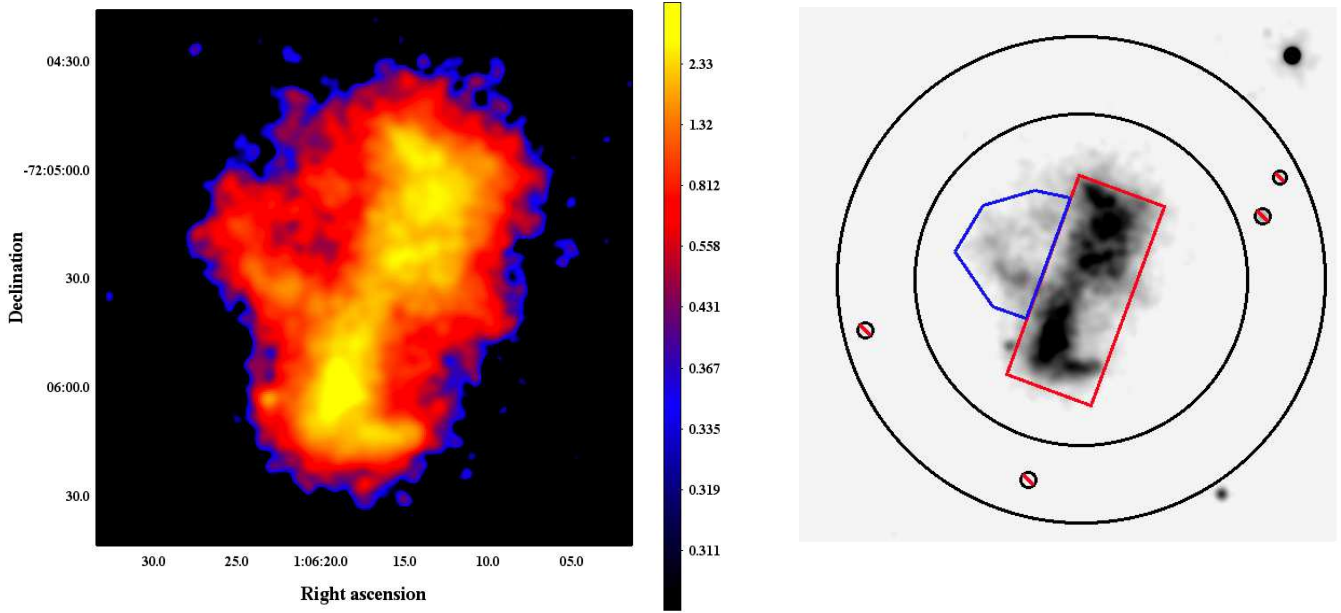}
\end{center}
\caption{{\it Left}: Full-band (0.5--8.0 keV) {\it Chandra} image of \source\ from the combined 528.6~ks data. Image has been smoothed with a 2.5\arcsec\ Gaussian kernel. {\it Right}: Full-band image with regions of spectral extraction overplotted: the ``bar'' and the ``arc'' regions in red and blue, respectively, and the region used for background subtraction (inside the black annulus, with the four point sources excluded). }
\label{fig:fullband}
\end{figure*}

We merged the imaging data using the {\sc ciao} command \hbox{\it reproject\_image\_grid}, which maps images into one reference frame. The resulting full-band (0.5--8.0 keV) {\it Chandra} image is shown in Figure~\ref{fig:fullband}. The net number of (background subtracted) full-band counts from the source in the merged observation is $\sim$3.1$\times10^{4}$ counts, sufficient to perform spatially-resolved spectroscopic analysis and to distinguish line emission in the spectra. For comparison, the 110~ks targeted ACIS-S3 observations (ObsIDs 9100 and 9810) from Lee et al. yielded $\sim$9500 net full-band counts (note that the count rates are different between observations depending on several factors, such as which ACIS chip imaged \source). We also produced narrow-band images at the energies of possible ejecta emission lines, but no unique structures/morphologies were evident, so we do not include them in this text. 

\section{Analyses and Results} \label{sec:results}

Using these archival data, we explore the explosive origin of \source, based on the X-ray morphology (Section~\ref{sec:morphology}) and on the ejecta metal abundances derived from X-ray spectral fits (Section~\ref{sec:spectra}).

\subsection{Morphology} \label{sec:morphology}

Recently, we have developed techniques to quantify the morphological properties of SNRs \citep{lopez09a}. We have applied these methods to archival {\it Chandra} and {\it Spitzer Space Telescope} images of young, ejecta-dominated SNRs ($\ls$25000 years old) to measure the asymmetry of their thermal X-ray and warm dust emission \citep{lopez09b,lopez11,peters13}. In these works, we found that the X-ray and IR morphology of SNRs can be used to differentiate Type Ia and CC SNRs. In particular, the thermal X-ray and IR emission of Type Ia SNRs are statistically more circular and mirror symmetric than that of CC SNRs. These results can be attributed to both the distinct geometries of the respective explosion mechanisms as well as the different circumstellar environments of Type Ia and CC SNe. 

Here, we apply the same symmetry technique as used previously, the power-ratio method (PRM), to the merged {\it Chandra} soft-band (0.5--2.1 keV) image of \source. Qualitatively, the morphology of \source\ appears bipolar in nature, but statistical comparison to a large sample of young SNRs reveals its distinctive nature. In this paper, we give a short overview of the PRM, and we refer the reader to our previous papers for the relevant background and equations. 

The PRM quantifies asymmetries via calculation of the multipole moments $\Psi_{\rm m}$ of emission in a circular aperture. It is derived similarly to the expansion of a two-dimensional gravitational potential, except an image's surface brightness replaces the mass surface density. The powers $P_{\rm m}$ are obtained by integrating the magnitude of each term $\Psi_{\rm m}$ over the aperture radius $R$. We divide the powers $P_{\rm m}$ by $P_{0}$ (the first term of the expansion) to normalize with respect to flux, and we set the origin position of our aperture to the centroid of the image so that the dipole power ratio $P_{1}/P_{0}$ approaches zero. In this case, the higher-order terms reflect the asymmetries at successively smaller scales. We focus on the two ratios in particular: the quadrupole power ratio $P_{2}/P_{0}$, which characterizes the ellipticity or elongation of a source, and the octupole power ratio $P_{3}/P_{0}$, which quantifies the mirror asymmetry of a source. 

We estimate the uncertainty in the power ratios of \source\ using a Monte Carlo approach. The soft-band image of \source\ was adaptively binned using the {\it AdaptiveBin} software \citep{sanders01} to smooth out noise. Subsequently, noise was added back by taking each adaptive pixel's intensity as the mean of a Poisson distribution and selecting a new intensity from that distribution. This process was repeated 100 times to produce 100 soft-band images of \source, and the 1$\sigma$ confidence limits on $P_{2}/P_{0}$ and $P_{3}/P_{0}$ are given by the sixteenth highest and lowest power ratios derived from the 100 images.

\begin{figure}
\begin{center}
\includegraphics[width=\columnwidth]{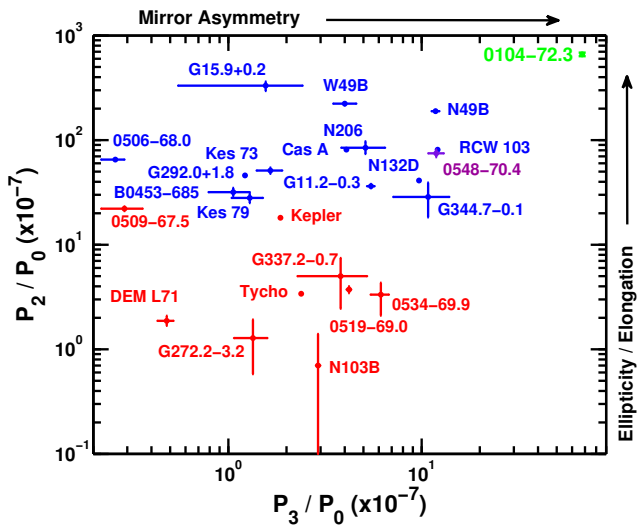}
\end{center}
\vspace{-5mm}
\caption{Results of the power-ratio method for \source\ (plotted in green) compared to 24 galactic and LMC SNRs (adapted from Figure~2 of \citealt{lopez11}). The quadrupole power ratio $P_{2}/P_{0}$ measures ellipticity/elongation; the octupole power ratio $P_{3}/P_{0}$ quantifies mirror asymmetry. Type Ia SNRs are plotted in red, and the CC SNRs are in blue (based on their ejecta abundance ratios). One source, SNR~0548$-$70.4, is in purple because of its anomalous abundance ratios. \source\ falls in the CC regime of the diagram. Furthermore, it is the most elliptical and mirror asymmetric of the sample, indicative of a complex environment and/or a highly bipolar explosion.}
\label{fig2}
\end{figure}

Figure~\ref{fig2} plots the resulting quadrupole power ratio $P_{2}/P_{0}$ versus octupole power ratio $P_{3}/P_{0}$ for \source\ compared to the values obtained for 24 galactic and LMC SNRs published in Figure~2 of \cite{lopez11}. \source\ falls in the CC SNR regime of the diagram, and it is the most elliptical and mirror asymmetric of all the sources, with $P_{2}/P_{0} = 660\pm26$ and $P_{3}/P_{0} = 67.7^{+3.8}_{-3.3}$. Thus, the X-ray morphology of \source\ is suggestive of a CC explosive origin. Moreover, recent hydrodynamical simulations have shown that the X-ray morphologies of jet-driven CC SNRs retain their bipolar structures over many hundreds of years \citep{diego13}. Thus, the extreme elongation and asymmetry of \source\ is consistent with a bipolar, jet-driven CC origin. 

\subsection{Spatially-Resolved Spectroscopy} \label{sec:spectra}

For our spectral analyses of \source, we extracted source and background spectra individually from each of the 56 ACIS observations using the {\sc ciao} command {\it specextract}. The regions of extraction are shown in Figure~\ref{fig:fullband} (right panel), including two source regions (the ``bar'' in red and the ``arc'' in blue) and an annulus region around \source\ (in black) used for the background. 

We began by replicating the modeling of \cite{lee11} on their two pointed ACIS-S observations of \source\ (ObsIDs 9100 and 9810). Due to an error in transcribing the SMC abundances from \cite{russell92}, Lee et al. adopted incorrect abundances in their fits, affecting their derived parameters (J.-J. Lee, private communication). Thus, we first verified that we could obtain the same results as \cite{lee11} using identical data and models with their (incorrect) adopted SMC abundances, and subsequently, we adopted the actual \cite{russell92} SMC abundances to see how the fits changed. All uncertainties listed below are 90\% confidence ranges for the given parameters. 

We extracted spectra from the same two regions as Lee et al. (the bar and the arc), and we modeled the spectra as a single absorbed non-equilibrium ionization (NEI) plasma using the plane-parallel shock model {\it vpshock} in XSPEC \citep{borkowski01}. We accounted for both the foreground absorption by the Milky Way, $N_{\rm H, MW}$, as well as the intrinsic absorption through the SMC,  $N_{\rm H, SMC}$, using the XSPEC model components {\it phabs} and {\it vphabs}, respectively. For the former component, we fixed the column density to $N_{\rm H, MW} = 2.2 \times10^{20}$ cm$^{-2}$, the H {\sc i} column density toward \source\ \citep{dickey90}. We began fitting with $N_{\rm H, SMC}$, the plasma temperature, and normalization thawed, and with all the metals of $N_{\rm H, SMC}$ and the NEI plasma frozen to SMC ISM abundances. As in Lee et al., these fits produced large residuals in the 0.8--1.2 keV range. In this band, Ne K and Fe L lines are prominent, so we subsequently freed both of these parameters and refit the data. 

Using this approach, we obtained best-fit parameters consistent within the uncertainties of the Lee et al. results for the bar and arc when using their SMC abundances. The results were similar whether we adopted \cite{anders89} or \cite{asplund09} for solar abundances. Therefore, we interpret all discrepancies from the Lee et al. results in the rest of this paper as stemming from the incorrectly adopted SMC abundances. In all analyses below, we opted to use \cite{asplund09} for our solar abundance values.  

We then remodeled the Lee et al. spectra but with the correct SMC abundances from \cite{russell92}\footnote{Relative to solar, the SMC abundances given in Table~1 of \cite{russell92} are as follows: He = 0.83; C = 0.15; N = 0.04; O = 0.13; Ne = 0.15; Mg = 0.25; Si = 0.30; S = 0.24; Ar = 0.18; Ca = 0.20; Fe = 0.15; Ni = 0.40.}. In both the bar and the arc, the resulting best-fit models had large ionization timescales ($\tau \gs10^{13}$ s cm$^{-3}$), which are long enough for the plasma to have reached collisional ionization equilibrium (CIE: \citealt{smith10}). Consequently, we substituted the variable abundance collisional ionization equilibrium plasma component {\it vapec} v2.0.2 \citep{smith01,foster12} for the {\it vpshock} component used above. Thus, written out in terms of XSPEC components, our model was \hbox{{\it phabs $\times$ vphabs $\times$ vapec}}. 

\begin{figure*}
\begin{center}
\includegraphics[width=\textwidth]{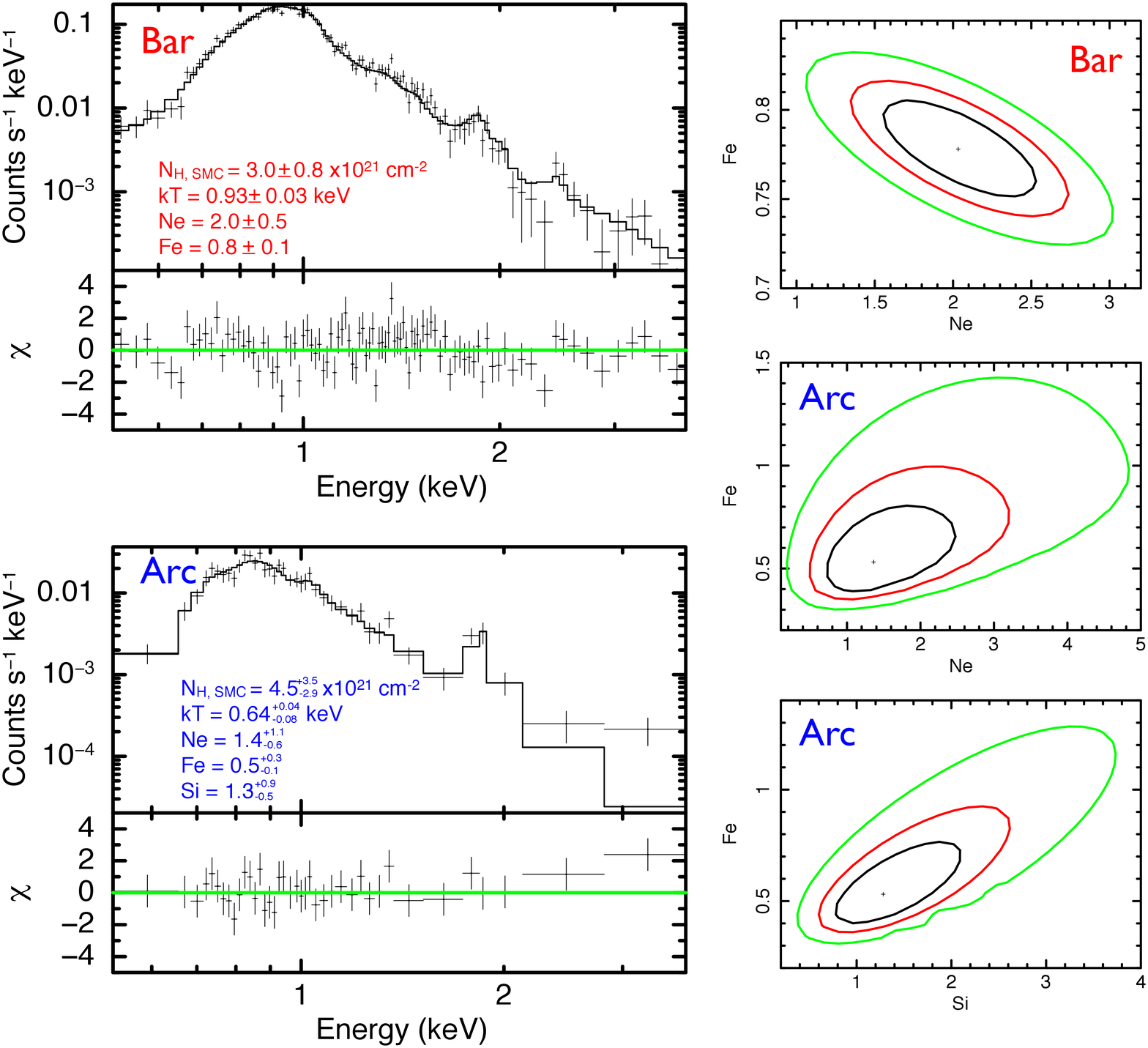}
\end{center}
\vspace{-5mm}
\caption{Spectral analyses of the bar and arc from the 110-ks pointed {\it Chandra} ACIS-S observations of \source. {\it Top left}: Spectra and model of bar region of the SNR. {\it Bottom left}: Same as top except for the bar region of \source. {\it Right panels}: 68\%, 95\%, and 99\% confidence contours of Fe versus Ne abundances in the bar (top), of Fe versus Ne abundances in the arc (middle), and of Fe vs Si abundances in the arc (bottom). All abundances are relative to solar by number. In the bar, we find enhanced abundances of Fe and Ne, and in the arc, we find enhanced abundances of Ne, Si, and Fe in the arc, although the confidence limits do not exclude the SMC ISM abundance values there.}
\label{fig3}
\end{figure*}

In the bar region, we began with all metals frozen to SMC ISM abundances; this model was statistically poor (with $\chi^{2} \approx$948 with 94 degrees of freedom) due to large residuals in the 0.8--1.2 keV range. Therefore, we freed the Ne and Fe abundances, and the resulting fit (shown in Figure~\ref{fig3}), improved dramatically, with $\chi^{2}$/d.o.f. $\approx$119/92. We followed a similar procedure with the Lee et al. data for the arc region. With all metals frozen to SMC values, we obtained a fit with $\chi^{2}$/d.o.f. $\approx$110/46. By thawing Ne and Fe and remodeling, the new fit improved statistically, with $\chi^{2}$/d.o.f. $\approx57$/44. Moreover, we thawed the Si abundance due to a large residual just below 2 keV, near the centroid of the Si {\sc xiii} line. This final fit, shown in Figure~\ref{fig3}, gave $\chi^{2}$/d.o.f. $\approx 46$/43, and all of the best-fit parameter values and their associated 1-$\sigma$ errors are listed in Table~\ref{table:spectraresults}. 

\begin{deluxetable*}{lcccccl}
\tablecolumns{7}\tablewidth{0pc} \tabletypesize{\footnotesize}
\tablewidth{0pt} \tablecaption{Spectral Results\tablenotemark{a}} 
\tablehead{\colhead{Region} & \colhead{$N_{\rm H, SMC}$\tablenotemark{b}} & \colhead{$kT$} & \colhead{Ne\tablenotemark{c}} & \colhead{Fe\tablenotemark{c}} & \colhead{Si\tablenotemark{c}} & \colhead{$\chi^{2}$/d.o.f.} \\
\colhead{} & \colhead{($\times10^{21}$ cm$^{-2}$)} & \colhead{(keV)} & \colhead{} & \colhead{} & \colhead{} & \colhead{}}
\startdata
\cutinhead{Models of Pointed ACIS Observations of \source}
Bar & 3.0$\pm$0.8 & 0.93$\pm$0.03 & 2.0$\pm$0.5 & 0.8$\pm$0.1 & 0.30 & 119/92 \\
Arc & 4.5$^{+3.5}_{-2.9}$ & 0.64$^{+0.04}_{-0.06}$ & 1.4$^{+0.1.1}_{-0.6}$ & 0.5$^{+0.3}_{-0.1}$ & 1.3$^{+0.9}_{-0.5}$ & 46/43 \\
\cutinhead{Models of All 56 ACIS Observations of \source}
Bar & 2.8$\pm$0.5 & 0.93$\pm$0.02 & 2.8$^{+0.7}_{-0.5}$ & 0.8$\pm$0.1 & 0.30 & 397/223 \\
Arc & 4.8$^{+3.0}_{-2.6}$ & 0.66$^{+0.04}_{-0.07}$ & 1.6$^{+0.5}_{-0.4}$ & 0.5$\pm$0.1 & 0.9$\pm$0.3 & 195/121
\enddata
\tablenotetext{a}{All error bars reflect 68\% uncertainty ranges.}
\tablenotetext{b}{Intrinsic absorption through the SMC. Foreground absorption by the Milky Way was fixed to $N_{\rm H, MW} = 2.2 \times 10^{20}$ cm$^{-2}$.} 
\tablenotetext{c}{Abundances are relative to solar by number. The Si abundance of 0.3 in the bar is the SMC ISM value.}
\label{table:spectraresults}
\end{deluxetable*}

We find two noteworthy differences, in addition to the long ionization timescales, from the fits reported by \cite{lee11}. First, we derive lower column densities, with $N_{\rm H, SMC} = (4.5^{+3.5}_{-2.9}) \times 10^{21}$ cm$^{-2}$ in the arc and $N_{\rm H, SMC} = (3.0\pm0.8) \times 10^{21}$ cm$^{-2}$ in the bar. These columns are slightly below the H {\sc i} column density measured toward \source\ of $\sim6 \times 10^{21}$ cm$^{-2}$ by \cite{stan99}, indicating the SNR may be on the nearer edge of the SMC. Secondly, we found that the arc may have enhanced abundances of Ne, Si, and Fe (as shown in Figure~\ref{fig3} and listed in Table~\ref{table:spectraresults}), with a Ne abundance of $1.4^{+1.1}_{-0.6}$, an Fe abundance of $0.5^{+0.3}_{-0.1}$, and a Si abundance of $1.3^{+0.9}_{-0.5}$ by number relative to solar. However, the confidence contours demonstrate that we cannot exclude the SMC ISM abundances (of 0.15 for both Ne and Fe and of 0.30 for Si) based on the 110 ks of pointed observation data. In the bar region, we also find statistically significantly enhanced abundances of Ne and Fe, with an Ne abundance of $2.0\pm0.5$ and an Fe abundance of $0.8\pm0.1$. These elevated abundances are required to adequately fit the spectrum, indicating an ejecta origin of the emission in the bar. 

\begin{figure*}
\begin{center}
\includegraphics[width=\textwidth]{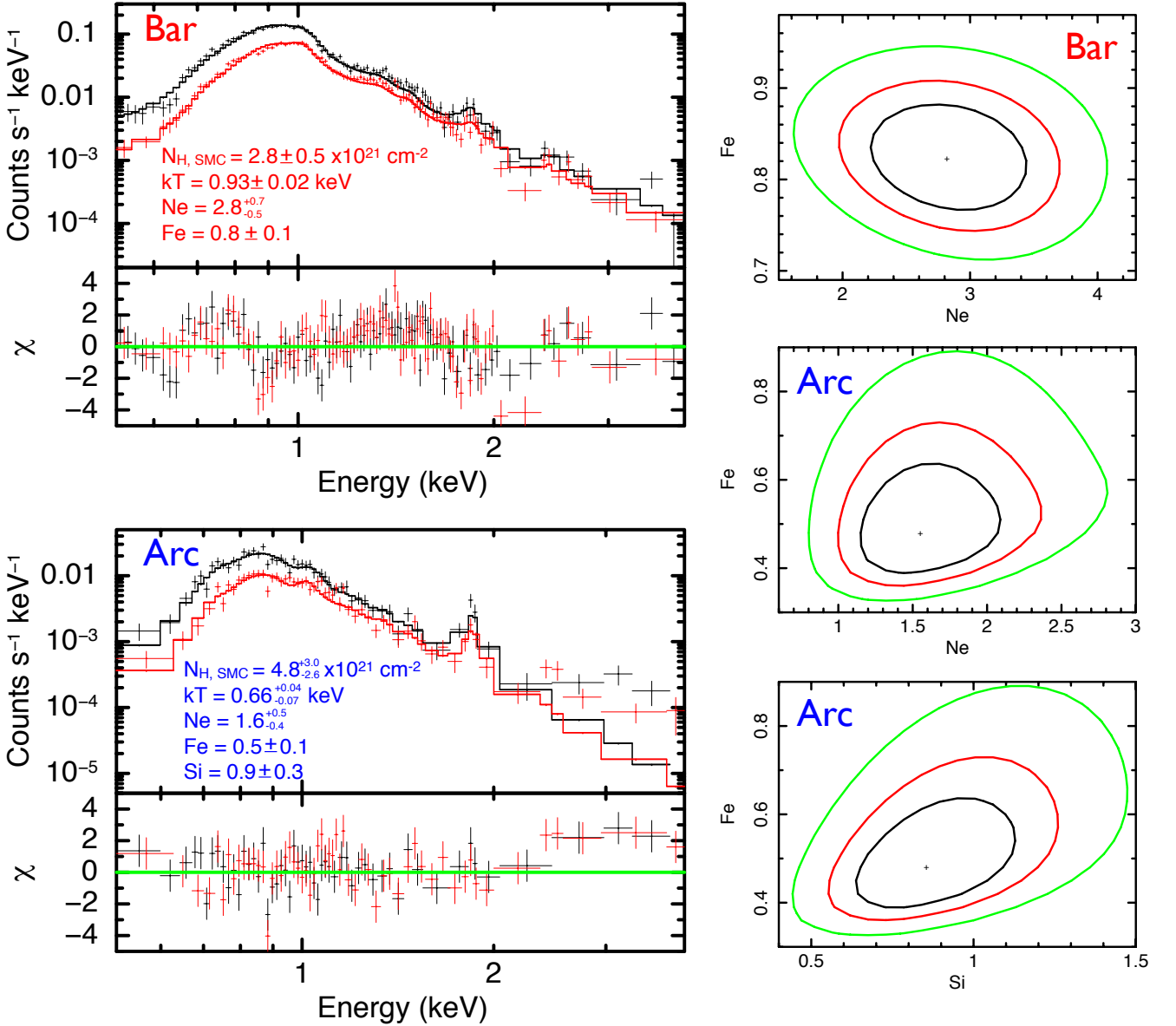}
\end{center}
\caption{Spectral analyses of the bar and arc from all 56 {\it Chandra} ACIS observations of \source. {\it Top left}: Back-illuminated (BI) and front-illuminated) spectra and model in black and red, respectively, of the bar region. {\it Bottom left}: Same as top except for the arc region of \source. {\it Right panels}: 68\%, 95\%, and 99\% confidence contours of Fe versus Ne abundances in the bar (top), of Fe versus Ne abundances in the arc (middle), and of Fe vs Si abundances in the arc (bottom). All abundances are relative to solar by number. In the bar, we find enhanced abundances of Fe and Ne, and in the arc, we find enhanced abundances of Ne, Si, and Fe. Using all 56 observations, we find that the abundances of these elements in the arc and bar are enhanced, indicating an ejecta origin.}
\label{fig4}
\end{figure*}

Subsequently, we modeled the background-subtracted spectra from the 56 observations to derive stricter limits on these parameters. Initially, we attempted to model the 56 individual spectra simultaneously of each region. However, this method did not yield reliable results, and the procedure often settled on parameters corresponding to local minima in chi-squared space, rather than the global minima. Therefore, we combined the spectra of the observations with \source\ on front-illuminated (FI) chips (ACIS I0--I3, S0, S2, S4, and S5; 44 observations) and those on back-illuminated (BI) chips (ACIS S1 and S3; 12 observations). We selected this procedure because the FI and BI chips have different responses: the BI chips have better chip-averaged energy resolution and responses which extend to lower energies than the FI chips. To combine the data, we used the {\sc ciao} command {\it combine\_spectra} to create FI and BI spectra from the arc and bar (see Figure~\ref{fig4}, left panels). Then, we modeled the FI and BI spectra simultaneously for each of the two regions. Following the presentation of our results below, we discuss the caveats associated with this approach. 

For the arc region, we began with the abundances frozen to their SMC values, and we obtained a fit with $\chi^{2}$/d.o.f. $\approx$369/124. We thawed the Ne and Fe abundances due to the large residuals from 0.8--1.2 keV, and the fit improved, with $\chi^{2}$/d.o.f. $\approx$213/122. Motivated by the emission peak near $\sim$1.8 keV, we thawed the Si abundance as well, and we obtained a fit with $\chi^{2}$/d.o.f. $\approx$195/121 (shown in Figure~\ref{fig4} and listed in Table~\ref{table:spectraresults}). This fit is statistically acceptable, but we note that the residuals above 2 keV are fairly large in the arc spectra. Due to the limited counts at those high energies, the nature of the excess flux is uncertain: it could be additional emission lines, a non-thermal component, or an additional hot thermal plasma component. 

In the bar, we followed a similar procedure; the fit with the abundances frozen to the SMC values was especially poor, with $\chi^{2}$/d.o.f. $\approx$ 3074/225. With Ne and Fe thawed, we obtained a fit with $\chi^{2}$/d.o.f. $\approx$ 397/223, as shown in Figure~\ref{fig4} and listed in Table~\ref{table:spectraresults}. The bar fit did not improve statistically with a thawed Si abundance, so we left it frozen to the SMC ISM abundance of 0.30 solar by number. 

The best-fit parameter values from the spectral modeling of the 56 observations (listed in Table~\ref{table:spectraresults}) are consistent within the uncertainties of those derived from just the 110~ks targeted observations (given in Figure~\ref{fig3}). The addition of the serendipitous data yields smaller error bars on all of the parameters, most notably on the Ne, Si, and Fe abundances. In the arc, the uncertainties are small enough to exclude an ISM origin with $>$99\% confidence for the Ne, Fe, and Si, suggesting these metals are ejecta material. In the bar, we also find statistically significant enhancement of Ne and Fe.  Furthermore, we investigated the O abundance in the bar and arc, but we are unable to set useful constraints or upper limits on the O abundance due to absorption toward \source. 

We note some caveats about the above spectral analyses. We have combined spectra from 56 observations using the {\sc ciao} command {\it combine\_spectra}, which co-adds imaging source and background spectra as well as instrument response files. To limit systematics, we have used the same source and background regions between observations. However, the different background rates between observations, dependent on which chip \source\ falls and the off-axis distance during each observation, causes some observations to be weighted more than others. In this procedure, the two ACIS-S3 observations totaling 110 ks would be weighted most heavily, yielding results similar to those derived by the two observations alone, but with smaller error bars. Furthermore, although simultaneous fitting of the individual 56 spectra would seem optimal, in practice the procedure gave unphysical results (e.g., temperatures $\sim$10 keV) as the fits found local minima of chi-squared space. 

\section{Discussion and Conclusions} \label{sec:conclusions}

\begin{figure*}
\begin{center}
\includegraphics[width=\textwidth]{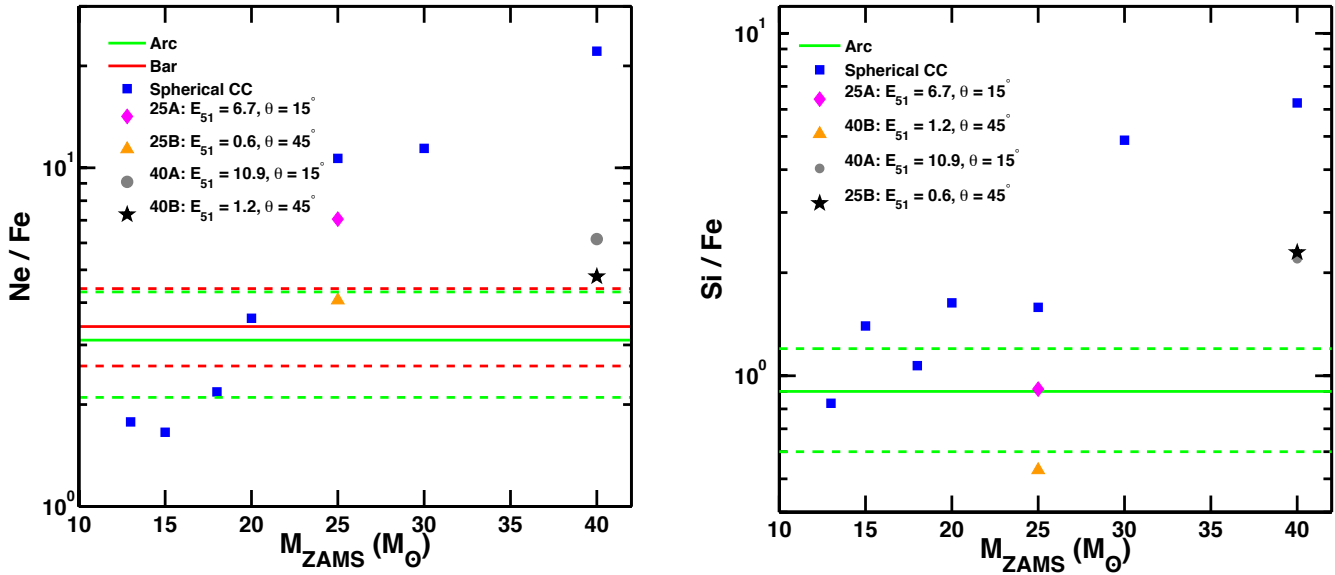}
\end{center}
\caption{Plots comparing the derived abundance ratios Ne/Fe (left) and Si/Fe (right) of \source\ to those predicted by CC SN models. Solid lines represent the best-fit values for the arc (light green) and the bar (red), with the dashed lines reflecting the 1-$\sigma$ uncertainties of those estimates. For comparison, we show the ratios produced by spherical CC models for different zero-age main-sequence mass stars $M_{\rm ZAMS}$ (blue squares, from \citealt{nomoto06}) and aspherical CC models of 25 $M_{\rm ZAMS}$ (purple diamond and black star) and 40 $M_{\rm ZAMS}$ (gray circle and stars from \citealt{maeda03}).}
\label{abundmodels}
\end{figure*}

In addition to the morphological evidence that \source\ was a jet-driven CC explosion, we can further constrain the explosive origin by comparing the ejecta abundances derived from the arc and bar to those predicted by SN models. Recall the above abundances are by number relative to solar. Converting the abundance ratios to be relative to mass and propagating the uncertainties, we have \hbox{Ne/Fe = 3.4$^{+1.0}_{-0.8}$} for the bar and \hbox{Ne/Fe = 3.1$^{+1.2}_{-1.0}$} and \hbox{Si/Fe = 0.9$\pm0.3$} for the arc. By comparison, Type Ia SNe produce much less Ne than we find in \source: the chemical yield of Type Ia SN models is Ne/Fe $\sim$0.006 in the deflagration case (model W7 of \citealt{nomoto84}) or Ne/Fe $\sim$0.001 in the delayed detonation case (model WDD2 of \citealt{nomoto97}). In contrast, the massive progenitors of CC SNe produce several orders of magnitude more Ne than Type Ia SNe, and thus they are more consistent with our measured ejecta abundances. In particular, the Ne/Fe ratio of the spherical CC models of \cite{nomoto06} increases with progenitor mass, from Ne/Fe $\sim$1.8 for a \hbox{13 $M_{\sun}$} progenitor to Ne/Fe $\sim$25 for a 40 $M_{\sun}$ progenitor (for a 1/5 solar metallicity star). Figure~\ref{abundmodels} shows graphically how our results compare to the values given by the models. The Ne/Fe ratios of the arc and bar of \source\ are most similar to the model predictions of the 18--20 $M_{\sun}$ progenitors, which have Ne/Fe $\sim$2.3 and $\sim$3.7, respectively. Our Si/Fe from the arc is also consistent with a 18--20 $M_{\sun}$ progenitor, as the spherical CC model gives Si/Fe $\sim$1.0--1.7. 

Given the elliptical/elongated morphology of \source, we also compare the abundance results to model predictions of aspherical CC SNe. \cite{maeda03} predict Ne/Fe ratios similar to those we obtain in the arc and bar, ranging from Ne/Fe$\sim$4--7 in their aspherical models of 25 $M_{\sun}$ and 40 $M_{\sun}$ progenitors (their models 25A, 25B, 40A, and 40B). However, the Si/Fe of the arc is below the yields of the aspherical 40 $M_{\sun}$ models (which give Si/Fe $\sim$2.1). Thus, our best-fit abundances are most consistent with either a spherical CC SN of a 18--20 $M_{\sun}$ progenitor or an aspherical CC SN of a 25 $M_{\sun}$ progenitor. Of these two possible progenitors, we favor an aspherical explosion scenario for \source\ due to its highly elliptical morphology. We also emphasize that hydrodynamical modeling of SNe has only sampled a small subset of the parameter space that can affect nucleosynthesis (e.g., jet opening angles, explosion energy, progenitor mass and metallicity). As such, our constraints on the progenitor mass from the metal abundances should be interpreted as approximate rather than as strict limits.

Here, we have assumed that all of the ejecta has been shock-heated and is well mixed, and thus, the best-fit abundances reflect the nucleosynthetic yield of the SN explosion. We can justify this assumption based on the age $t$ of \source, using the Sedov-Taylor solution: 

\begin{equation}
t = 17.3~ \bigg( \frac{10^{51}~{\rm erg}}{E_{\rm SN}} \bigg)^{1/2} \bigg( \frac{n_{0}}{1.0~{\rm cm}^{-3}} \bigg)^{1/2} \bigg( \frac{R_{\rm s}}{1~{\rm pc}} \bigg)^{5/2}~~{\rm years}.
\end{equation}

\noindent
The diameter of \source\ is $\sim$1.8\arcmin, corresponding to $R_{\rm s} \sim$16 pc, assuming a distance of 61 kpc to the SMC \citep{hilditch05}. Given this radius, the approximate age of \source\ is $t \sim$ 17600 years, for an explosion of energy $E_{\rm SN} = 10^{51}$ erg in an ambient electron density of $n_{0} = 1.0$ cm$^{-3}$. At this relatively evolved stage, the reverse shock would have completed its propagation inward relative to the ejecta (see \citealt{truelove99}), heating the ejecta to X-ray emitting temperatures. 

In this paper, we have employed 528.6 ks of serendipitous {\it Chandra} observations to reconsider the explosive origin of the SMC \source. We have demonstrated that the X-ray morphology of \source\ is extremely elliptical/elongated compared to other young SNRs in the MW and the LMC, suggesting it is the result of a core-collapse SN. Furthermore, we have performed spectral modeling of two regions of \source, and we have found enhanced abundances of Ne, Si, and Fe. Through comparison to the nucleosynthesis predictions of Type Ia and CC SN models, we demonstrate that the yields are consistent with either a spherical CC SN of a 18--20 $M_{\sun}$ progenitor or an aspherical CC  SN of a 25 $M_{\sun}$ progenitor. Given the bipolar morphology of the SNR, we consider the latter scenario to be more likely, in which case it would be the second SNR likely to have had a bipolar origin (the first being W49B in the MW: \citealt{lopez13}). 

\begin{figure}
\begin{center}
\includegraphics[width=\columnwidth]{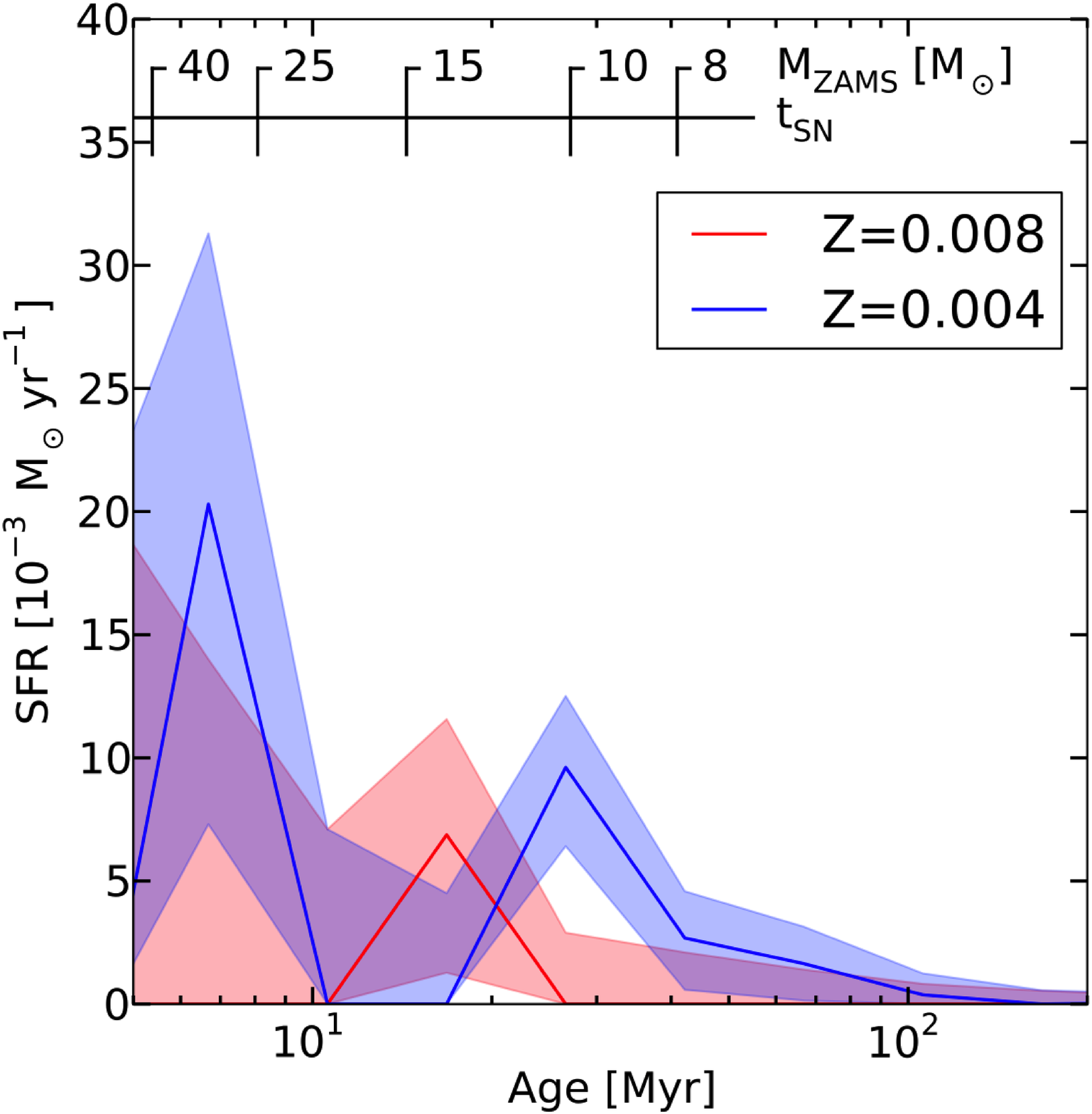}
\end{center}
\caption{Star formation rate (SFR) versus look-back time for two metallicity stellar populations ($Z = 0.004$ and $Z=0.008$) at the site of \source, taken from the star-formation history (SFH) maps of the SMC produced by \cite{harris04}. The solid lines represent the best-fit SFR reported by \cite{harris04}, and the shaded regions give their 68\% confidence limits. At the top of the figure, we mark the time $t_{\rm SN}$ when the progenitor star of a given zero-age main sequence mass $M_{\rm ZAMS}$ would have formed (using the single star models of \citealt{eldridge08}). We find that the peak SFR occurred at $\sim$7 Myr ago for a metallicity $Z = 0.004$ or at $\sim$15 Myr ago for $Z = 0.008$. These ages correspond to progenitor stars of mass $M_{\rm ZAMS} \sim 30 M_{\sun}$ or $M_{\rm ZAMS} \sim 13 M_{\sun}$, respectively. Thus, the SFH at the site of \source\ is also suggestive of a CC origin of this SNR. }
\label{fig5}
\end{figure}

As a consistency check on the nature of \source, we investigate the star formation history (SFH) at the site of \source\ using the spatially-resolved SFH maps of the SMC produced by \cite{harris04}. These authors obtained {\it UBVI} photometry of 6 million stars across the SMC and produced a uniform grid of 12\arcmin $\times$12\arcmin\ regions, each with its own SFH derived from StarFISH \citep{harris01}. Figure~\ref{fig5} shows a plot of the star formation rate (SFR) versus age obtained by Harris \& Zaritsky at the location of \source. The peak SFR occurred at $\sim$7 Myr ago for a metallicity $Z = 0.004$ or at $\sim$15 Myr ago for $Z = 0.008$. These ages correspond to stars of zero-age main sequence mass $M_{\rm ZAMS} \sim 30 M_{\sun}$ or $M_{\rm ZAMS} \sim 13 M_{\sun}$, respectively, using the single star models of \cite{eldridge08}. Thus, the SFH at the site of \source\ -- particularly that of the low-metallicity case -- gives the appropriate age for a massive star ($\sim$25--40 $M_{\sun}$) origin of \source. Although the averaged SFHs derived from resolved stellar populations can be misleading regarding SN progenitors (especially for Type Ia SNRs; see the discussion of \citealt{bad09}), this approach provides a complementary means to verify our results.

The identification of a bipolar SNR in the SMC is remarkable given that only $\sim$25 SNRs are known in the SMC \citep{bad10}. By comparison, only one SNR in the MW has been shown to be bipolar to date \citep{lopez13} out of $\sim$300 SNRs \citep{green09}, an order of magnitude larger sample. Yet if the SN rate is proportional to the star formation rate (SFR) of a galaxy, then the number of bipolar SNe in the MW should be a factor of 20 {\it greater} than the SMC (assuming a SFR of $\sim$1 $M_{\sun}$ year$^{-1}$ in the MW and of $\sim$0.05 $M_{\sun}$ year$^{-1}$ in the SMC; \citealt{wilke04,rob10}). Thus, a bipolar SNR in the SMC may suggest that jet-driven explosions may occur more commonly in the SMC, a low metallicity galaxy (with $\sim$0.2 $Z_{\sun}$: \citealt{russell92}). If bipolar SNe happened at the same rate in the MW and SMC, we would expect to have $\sim$12 bipolar SNRs out of the 300 in our galaxy. While no other MW SNRs currently show evidence of bipolarity (based on e.g., morphology, abundances, and star-formation histories: \citealt{lopez11}), a full statistical analysis is necessary to demonstrate definitively that bipolar SNe are more likely in the SMC compared to the MW. 

If this result holds, it is consistent with recent work showing bipolar SNe prefer low-metallicity environments (e.g., \citealt{ramirezruiz02,izzard04,fruchter06}), since rapid rotation without extensive mass loss is required to produce these explosions. In particular, broad-line (BL) Type Ic SNe (which are thought to be jet-driven explosions) are observed to occur at lower median metallicity than typical Type Ib/c SNe \citep{sanders12}. Furthermore, at sufficiently low metallicity ($\sim$0.2--0.6 $Z_{\sun}$), BL-Ic SNe are associated with long-duration GRBs \citep{kaneko07,modjaz08}, and 75\% of long GRBs are located in host galaxies with $\ls$0.5 $Z_{\sun}$ \citep{graham13}.  Our findings would support this framework of jet-driven explosions happening more frequently at low metallicity. 

In the future, Astro-H observations of \source\ will be able to differentiate the Ne and Fe L emission lines that comprise the bump in the X-ray spectra at 0.9--1.1 keV, facilitating tighter constraints on the relative abundances of the two elements. Although the distance to the SMC would preclude the X-ray detection of a thermal emission from any central compact object, non-thermal emission or pulsations may be observable if the SNR has a pulsar. 

\acknowledgements

We acknowledge helpful discussions with J.-J. Lee in writing this paper. Support for LAL was provided by NASA through the Einstein Fellowship Program, grant PF1--120085, and the MIT Pappalardo Fellowship in Physics. ERR acknowledges support from the David and Lucile Packard Foundation and NSF grant AST--0847563. DC and POS acknowledge support for this work provided by NASA through the Smithsonian Astrophysical Observatory contract SV3--73016 to MIT for support of the Chandra X-ray Center, which is operated by the Smithsonian Astrophysical Observatory for and on behalf of NASA under contract NAS8--03060.


\begin{thebibliography}{51}
\expandafter\ifx\csname natexlab\endcsname\relax\def\natexlab#1{#1}\fi

\bibitem[{{Anders} \& {Grevesse}(1989)}]{anders89}
{Anders}, E., \& {Grevesse}, N. 1989, \gca, 53, 197

\bibitem[{{Arnaud}(1996)}]{arnaud96}
{Arnaud}, K.~A. 1996, in Astronomical Society of the Pacific Conference Series,
  Vol. 101, Astronomical Data Analysis Software and Systems V, ed. G.~H.
  {Jacoby} \& J.~{Barnes}, 17

\bibitem[{{Asplund} {et~al.}(2009){Asplund}, {Grevesse}, {Sauval}, \&
  {Scott}}]{asplund09}
{Asplund}, M., {Grevesse}, N., {Sauval}, A.~J., \& {Scott}, P. 2009, \araa, 47,
  481

\bibitem[{{Aubourg} {et~al.}(2008){Aubourg}, {Tojeiro}, {Jimenez}, {Heavens},
  {Strauss}, \& {Spergel}}]{aubourg08}
{Aubourg}, {\'E}., {Tojeiro}, R., {Jimenez}, R., {Heavens}, A., {Strauss},
  M.~A., \& {Spergel}, D.~N. 2008, \aap, 492, 631

\bibitem[{{Badenes} {et~al.}(2007){Badenes}, {Hughes}, {Bravo}, \&
  {Langer}}]{bad07}
{Badenes}, C., {Hughes}, J.~P., {Bravo}, E., \& {Langer}, N. 2007, \apj, 662,
  472

\bibitem[{{Badenes} {et~al.}(2009){Badenes}, {Harris}, {Zaritsky}, \&
  {Prieto}}]{bad09} {Badenes}, C., {Harris}, 
J., {Zaritsky}, D., \& {Prieto}, J.~L.\ 2009, \apj, 700, 727 

\bibitem[{{Badenes} {et~al.}(2010){Badenes}, {Maoz}, \& {Draine}}]{bad10}
{Badenes}, C., {Maoz}, D., \& {Draine}, B.~T. 2010, \mnras, 407, 1301

\bibitem[{{Borkowski} {et~al.}(2001){Borkowski}, {Lyerly}, \&
  {Reynolds}}]{borkowski01}
{Borkowski}, K.~J., {Lyerly}, W.~J., \& {Reynolds}, S.~P. 2001, \apj, 548, 820

\bibitem[{{Dan} {et~al.}(2012){Dan}, {Rosswog}, {Guillochon}, \&
  {Ramirez-Ruiz}}]{dan12}
{Dan}, M., {Rosswog}, S., {Guillochon}, J., \& {Ramirez-Ruiz}, E. 2012, \mnras,
  422, 2417

\bibitem[{{Dickey} \& {Lockman}(1990)}]{dickey90}
{Dickey}, J.~M., \& {Lockman}, F.~J. 1990, \araa, 28, 215

\bibitem[{{Eldridge} et al.(2008)}]{eldridge08} {Eldridge}, J.~J., 
{Izzard}, R.~G., \& {Tout}, C.~A.\ 2008, \mnras, 384, 1109 

\bibitem[{{Foster} {et~al.}(2012){Foster}, {Ji}, {Smith}, \&
  {Brickhouse}}]{foster12}
{Foster}, A.~R., {Ji}, L., {Smith}, R.~K., \& {Brickhouse}, N.~S. 2012, \apj,
  756, 128

\bibitem[{{Fruchter} {et~al.}(2006){Fruchter}, {Levan}, {Strolger},
  {Vreeswijk}, {Thorsett}, {Bersier}, {Burud}, {Castro Cer{\'o}n},
  {Castro-Tirado}, {Conselice}, {Dahlen}, {Ferguson}, {Fynbo}, {Garnavich},
  {Gibbons}, {Gorosabel}, {Gull}, {Hjorth}, {Holland}, {Kouveliotou}, {Levay},
  {Livio}, {Metzger}, {Nugent}, {Petro}, {Pian}, {Rhoads}, {Riess}, {Sahu},
  {Smette}, {Tanvir}, {Wijers}, \& {Woosley}}]{fruchter06}
{Fruchter}, A.~S., {et~al.} 2006, \nat, 441, 463

\bibitem[{{Gehrels} {et~al.}(2009){Gehrels}, {Ramirez-Ruiz}, \&
  {Fox}}]{gehrels09}
{Gehrels}, N., {Ramirez-Ruiz}, E., \& {Fox}, D.~B. 2009, \araa, 47, 567

\bibitem[{{Gonz{\'a}lez-Casanova} {et~al.}(2014){Gonz{\'a}lez-Casanova}, {De Colle},
  {Ramirez-Ruiz}, \& {Lopez}}]{diego13}
{Gonz{\'a}lez-Casanova}, D.~F., {De Colle}, F., {Ramirez-Ruiz}, E., \& {Lopez},
  L.~A.\ 2014, \apjl, 781, L26 

\bibitem[{{Graham} \& {Fruchter}(2013)}]{graham13}
{Graham}, J.~F., \& {Fruchter}, A.~S. 2013, \apj, 774, 119

\bibitem[{{Green}(2009)}]{green09}
{Green}, D.~A. 2009, Bulletin of the Astronomical Society of India, 37, 45

\bibitem[Guillochon et al.(2010)]{james} Guillochon, J., Dan, 
M., Ramirez-Ruiz, E., \& Rosswog, S.\ 2010, \apjl, 709, L64

\bibitem[{{Harris} \& {Zaritsky}(2001)}]{harris01} {Harris}, J., \& {Zaritsky}, D.\ 2001, \apjs, 136, 25 

\bibitem[{{Harris} \& {Zaritsky}(2004)}]{harris04} {Harris}, J., \& {Zaritsky}, D.\ 2004, \aj, 127, 1531 

\bibitem[{{Hilditch} {et~al.}(2005){Hilditch}, {Howarth}, \& {Harries}}]{hilditch05}
{Hilditch}, R.~W., {Howarth}, I.~D., \& {Harries}, T.~J. 2005, \mnras, 357, 304

\bibitem[{{Iben} \& {Tutukov}(1984)}]{iben84}
{Iben}, Jr., I., \& {Tutukov}, A.~V. 1984, \apjs, 54, 335

\bibitem[{{Izzard} {et~al.}(2004){Izzard}, {Ramirez-Ruiz}, \&
  {Tout}}]{izzard04}
{Izzard}, R.~G., {Ramirez-Ruiz}, E., \& {Tout}, C.~A. 2004, \mnras, 348, 1215

\bibitem[Kaneko et al.(2007)]{kaneko07} Kaneko, Y., 
Ramirez-Ruiz, E., Granot, J., et al.\ 2007, \apj, 654, 385

\bibitem[{{Kerzendorf} {et~al.}(2013){Kerzendorf}, {Childress},
  {Scharwaechter}, {Do}, \& {Schmidt}}]{kerz13}
{Kerzendorf}, W.~E., {Childress}, M., {Scharwaechter}, J., {Do}, T., \&
  {Schmidt}, B.~P. 2013, ArXiv e-prints

\bibitem[{{Koo} {et~al.}(2007){Koo}, {Lee}, {Moon}, {Lee}, {Seok}, {Lee},
  {Hong}, {Lee}, {Kaneda}, {Ita}, {Jeong}, {Onaka}, {Sakon}, {Nakagawa}, \&
  {Murakami}}]{koo07}
{Koo}, B.-C., {et~al.} 2007, \pasj, 59, 455

\bibitem[{{Law} {et~al.}(2009){Law}, {Kulkarni}, {Dekany}, {Ofek}, {Quimby},
  {Nugent}, {Surace}, {Grillmair}, {Bloom}, {Kasliwal}, {Bildsten}, {Brown},
  {Cenko}, {Ciardi}, {Croner}, {Djorgovski}, {van Eyken}, {Filippenko}, {Fox},
  {Gal-Yam}, {Hale}, {Hamam}, {Helou}, {Henning}, {Howell}, {Jacobsen},
  {Laher}, {Mattingly}, {McKenna}, {Pickles}, {Poznanski}, {Rahmer}, {Rau},
  {Rosing}, {Shara}, {Smith}, {Starr}, {Sullivan}, {Velur}, {Walters}, \&
  {Zolkower}}]{law05}
{Law}, N.~M., {et~al.} 2009, \pasp, 121, 1395

\bibitem[{{Leaman} {et~al.}(2011){Leaman}, {Li}, {Chornock}, \&
  {Filippenko}}]{leaman11}
{Leaman}, J., {Li}, W., {Chornock}, R., \& {Filippenko}, A.~V. 2011, \mnras,
  412, 1419

\bibitem[{{Lee} {et~al.}(2011){Lee}, {Park}, {Hughes}, {Slane}, \&
  {Burrows}}]{lee11}
{Lee}, J.-J., {Park}, S., {Hughes}, J.~P., {Slane}, P.~O., \& {Burrows}, D.~N.
  2011, \apjl, 731, L8

\bibitem[{{Lopez} {et~al.}(2013{\natexlab{a}}){Lopez}, {Pearson},
  {Ramirez-Ruiz}, {Castro}, {Yamaguchi}, {Slane}, \& {Smith}}]{lopez13b}
{Lopez}, L.~A., {Pearson}, S., {Ramirez-Ruiz}, E., {Castro}, D., {Yamaguchi},
  H., {Slane}, P.~O., \& {Smith}, R.~K. 2013{\natexlab{a}}, ArXiv e-prints

\bibitem[{{Lopez} {et~al.}(2009{\natexlab{a}}){Lopez}, {Ramirez-Ruiz},
  {Badenes}, {Huppenkothen}, {Jeltema}, \& {Pooley}}]{lopez09b}
{Lopez}, L.~A., {Ramirez-Ruiz}, E., {Badenes}, C., {Huppenkothen}, D.,
  {Jeltema}, T.~E., \& {Pooley}, D.~A. 2009{\natexlab{a}}, \apjl, 706, L106

\bibitem[{{Lopez} {et~al.}(2013{\natexlab{b}}){Lopez}, {Ramirez-Ruiz},
  {Castro}, \& {Pearson}}]{lopez13}
{Lopez}, L.~A., {Ramirez-Ruiz}, E., {Castro}, D., \& {Pearson}, S.
  2013{\natexlab{b}}, \apj, 764, 50

\bibitem[{{Lopez} {et~al.}(2011){Lopez}, {Ramirez-Ruiz}, {Huppenkothen},
  {Badenes}, \& {Pooley}}]{lopez11}
{Lopez}, L.~A., {Ramirez-Ruiz}, E., {Huppenkothen}, D., {Badenes}, C., \&
  {Pooley}, D.~A. 2011, \apj, 732, 114

\bibitem[{{Lopez} {et~al.}(2009{\natexlab{b}}){Lopez}, {Ramirez-Ruiz},
  {Pooley}, \& {Jeltema}}]{lopez09a}
{Lopez}, L.~A., {Ramirez-Ruiz}, E., {Pooley}, D.~A., \& {Jeltema}, T.~E.
  2009{\natexlab{b}}, \apj, 691, 875

\bibitem[{{Maeda} \& {Nomoto}(2003)}]{maeda03}
{Maeda}, K., \& {Nomoto}, K. 2003, \apj, 598, 1163

\bibitem[{{Modjaz} {et~al.}(2008){Modjaz}, {Kewley}, {Kirshner}, {Stanek},
  {Challis}, {Garnavich}, {Greene}, {Kelly}, \& {Prieto}}]{modjaz08}
{Modjaz}, M., {et~al.} 2008, \aj, 135, 1136

\bibitem[{{Nomoto} {et~al.}(1997){Nomoto}, {Iwamoto}, {Nakasato}, {Thielemann},
  {Brachwitz}, {Tsujimoto}, {Kubo}, \& {Kishimoto}}]{nomoto97}
{Nomoto}, K., {Iwamoto}, K., {Nakasato}, N., {Thielemann}, F.-K., {Brachwitz},
  F., {Tsujimoto}, T., {Kubo}, Y., \& {Kishimoto}, N. 1997, Nuclear Physics A,
  621, 467

\bibitem[{{Nomoto} {et~al.}(1984){Nomoto}, {Thielemann}, \& {Yokoi}}]{nomoto84}
{Nomoto}, K., {Thielemann}, F.-K., \& {Yokoi}, K. 1984, \apj, 286, 644

\bibitem[{{Nomoto} {et~al.}(2006){Nomoto}, {Tominaga}, {Umeda}, {Kobayashi}, \&
  {Maeda}}]{nomoto06}
{Nomoto}, K., {Tominaga}, N., {Umeda}, H., {Kobayashi}, C., \& {Maeda}, K.
  2006, Nuclear Physics A, 777, 424

\bibitem[{{Pakmor} {et~al.}(2012){Pakmor}, {Kromer}, {Taubenberger}, {Sim},
  {R{\"o}pke}, \& {Hillebrandt}}]{pakmor12}
{Pakmor}, R., {Kromer}, M., {Taubenberger}, S., {Sim}, S.~A., {R{\"o}pke},
  F.~K., \& {Hillebrandt}, W. 2012, \apjl, 747, L10

\bibitem[{{Peters} {et~al.}(2013){Peters}, {Lopez}, {Ramirez-Ruiz}, {Stassun},
  \& {Figueroa-Feliciano}}]{peters13}
{Peters}, C.~L., {Lopez}, L.~A., {Ramirez-Ruiz}, E., {Stassun}, K.~G., \&
  {Figueroa-Feliciano}, E. 2013, \apjl, 771, L38

\bibitem[{{Podsiadlowski} {et~al.}(2004){Podsiadlowski}, {Mazzali}, {Nomoto},
  {Lazzati}, \& {Cappellaro}}]{pod04}
{Podsiadlowski}, P., {Mazzali}, P.~A., {Nomoto}, K., {Lazzati}, D., \&
  {Cappellaro}, E. 2004, \apjl, 607, L17

\bibitem[Ramirez-Ruiz et al.(2002)]{ramirezruiz02} Ramirez-Ruiz, E., Lazzati, D., \& Blain, A.~W.\ 2002, \apjl, 565, L9 

\bibitem[{{Ramirez-Ruiz} \& {MacFadyen}(2010)}]{ramirezruiz10}
{Ramirez-Ruiz}, E., \& {MacFadyen}, A.~I. 2010, \apj, 716, 1028

\bibitem[{{Robitaille} \& {Whitney}(2010)}]{rob10}
{Robitaille}, T.~P., \& {Whitney}, B.~A. 2010, \apjl, 710, L11

\bibitem[{{Russell} \& {Dopita}(1992)}]{russell92}
{Russell}, S.~C., \& {Dopita}, M.~A. 1992, \apj, 384, 508

\bibitem[{{Sanders} \& {Fabian}(2001)}]{sanders01}
{Sanders}, J.~S., \& {Fabian}, A.~C. 2001, \mnras, 325, 178

\bibitem[{{Sanders} {et~al.}(2012){Sanders}, {Soderberg}, {Levesque}, {Foley},
  {Chornock}, {Milisavljevic}, {Margutti}, {Berger}, {Drout}, {Czekala}, \&
  {Dittmann}}]{sanders12}
{Sanders}, N.~E., {et~al.} 2012, \apj, 758, 132

\bibitem[{{Scannapieco} \& {Bildsten}(2005)}]{scann05}
{Scannapieco}, E., \& {Bildsten}, L. 2005, \apjl, 629, L85

\bibitem[{{Schaefer} \& {Pagnotta}(2012)}]{schaefer12}
{Schaefer}, B.~E., \& {Pagnotta}, A. 2012, \nat, 481, 164

\bibitem[{{Sedov}(1959)}]{sedov59} {Sedov}, L.~I.\ 1959, Similarity 
and Dimensional Methods in Mechanics, New York: Academic Press, 1959

\bibitem[{{Smith} {et~al.}(2001){Smith}, {Brickhouse}, {Liedahl}, \&
  {Raymond}}]{smith01}
{Smith}, R.~K., {Brickhouse}, N.~S., {Liedahl}, D.~A., \& {Raymond}, J.~C.
  2001, \apjl, 556, L91

\bibitem[{{Smith} \& {Hughes}(2010)}]{smith10}
{Smith}, R.~K., \& {Hughes}, J.~P. 2010, \apj, 718, 583

\bibitem[{{Soderberg} {et~al.}(2010){Soderberg}, {Chakraborti}, {Pignata},
  {Chevalier}, {Chandra}, {Ray}, {Wieringa}, {Copete}, {Chaplin},
  {Connaughton}, {Barthelmy}, {Bietenholz}, {Chugai}, {Stritzinger}, {Hamuy},
  {Fransson}, {Fox}, {Levesque}, {Grindlay}, {Challis}, {Foley}, {Kirshner},
  {Milne}, \& {Torres}}]{soderberg10}
{Soderberg}, A.~M., {et~al.} 2010, \nat, 463, 513

\bibitem[{{Stanimirovic} {et~al.}(1999){Stanimirovic}, {Staveley-Smith},
  {Dickey}, {Sault}, \& {Snowden}}]{stan99}
{Stanimirovic}, S., {Staveley-Smith}, L., {Dickey}, J.~M., {Sault}, R.~J., \&
  {Snowden}, S.~L. 1999, \mnras, 302, 417

\bibitem[{{Truelove} \& {McKee}(1999)}]{truelove99} {Truelove}, J.~K., \& {McKee}, C.~F.\ 1999, \apjs, 120, 299 

\bibitem[{{Vink}(2012)}]{vink12}
{Vink}, J. 2012, \aapr, 20, 49

\bibitem[{{Webbink}(1984)}]{webbink84}
{Webbink}, R.~F. 1984, \apj, 277, 355

\bibitem[{{Whelan} \& {Iben}(1973)}]{whelan73}
{Whelan}, J., \& {Iben}, Jr., I. 1973, \apj, 186, 1007

\bibitem[{{Wilke} {et~al.}(2004){Wilke}, {Klaas}, {Lemke}, {Mattila},
  {Stickel}, \& {Haas}}]{wilke04}
{Wilke}, K., {Klaas}, U., {Lemke}, D., {Mattila}, K., {Stickel}, M., \& {Haas},
  M. 2004, \aap, 414, 69

\bibitem[{{Williams} {et~al.}(2011){Williams}, {Blair}, {Blondin}, {Borkowski},
  {Ghavamian}, {Long}, {Raymond}, {Reynolds}, {Rho}, \& {Winkler}}]{williams11}
{Williams}, B.~J., {et~al.} 2011, \apj, 741, 96

\end{thebibliography}
\end{document}